\documentclass[article,nojss]{jss}

\usepackage{amsmath}
\usepackage{amssymb}
\usepackage{bbold}

\newcommand{\betac}{\beta} %symbol for regression parameter
\newcommand{\betad}{d} %dimension of regression coefficient
\newcommand{\betav}{\boldsymbol{\betac}}

\newcommand{\bfz}{{\mathbf{0}}}

\newcommand{\Bino}[1]{\mbox{\rm BiNom}\left(#1\right)}
\newcommand{\errordiff}{\varepsilon}

\newcommand{\Ferror}{F_{\errordiff}}
\newcommand{\ferror}{f_{\errordiff}}

\newcommand{\GenLogistic}[2]{\mathcal{GL}_{\mbox{\tiny #1}}\left(#2\right)}

\newcommand{\indic}[1]{I\{#1\}}

\newcommand{\labset}{L}

\newcommand{\Logistic}{\mathcal{LO}}
\newcommand{\logit}{\mbox{\rm logit}}

\newcommand{\Normalpdf}{\phi}
\newcommand{\Normalcdf}{\Phi}

\newcommand{\omegaH}{\omega}

\newcommand{\PG}[1]{\mathcal{PG}\left(#1\right)}

\newcommand{\pl}{\pi}  %notation used for the probabilities in a logit model
\newcommand{\Probsym}{\mbox{\rm Pr}}
\newcommand{\Xbeta}{{\mathbf \Xz}}

\newcommand{\Xz}{x}% symbol used for the row in a regressor matrix.

\newcommand{\yc}{y}
\newcommand{\ym}{{\mathbf \yc}}
\newcommand{\ypro}{{u}}
\newcommand{\yprodiff}{z}

\newcommand{\yprow}{{w}}

\newcommand{\yprov}{{v}}

\newcommand{\Probab}[1]{\Probsym (#1)}

\author{Gregor Zens\\Vienna University of\\Economics and Business \And 
        Sylvia Fr{\"u}hwirth-Schnatter\\Vienna University of\\Economics and Business \And
        Helga Wagner\\Johannes Kepler\\University Linz}
\title{Efficient Bayesian Modeling of Binary and Categorical Data in \proglang{R}: The \pkg{UPG} Package}

\Plainauthor{Gregor Zens, Sylvia Fr{\"u}hwirth-Schnatter, Helga Wagner} %% comma-separated
\Plaintitle{Efficient Bayesian Modeling of Binary and Categorical Data in R: The UPG Package} %% without formatting
\Shorttitle{\pkg{UPG}: Efficient Bayesian Models for Binary and Categorical data} %% a short title (if necessary)

\Abstract{
In this vignette, we introduce the \pkg{UPG} package for efficient Bayesian inference in probit, logit, multinomial logit and binomial logit models. \pkg{UPG} offers a convenient estimation framework for balanced and imbalanced data settings where sampling efficiency is ensured through marginal data augmentation. \pkg{UPG} provides several methods for fast production of output tables and summary plots that are easily accessible to a broad range of users.}
\Plainkeywords{logit, multinomial, probit, binomial, imbalanced data, MCMC, data augmentation} %% without formatting
\Address{
  Gregor Zens\\
  Vienna University of Economics and Business\\
  1020 Vienna, Austria\\
  E-mail: \email{gzens@wu.ac.at}\\
  URL: \url{gregorzens.github.io}
}
\begin{document}

%% include your article here, just as usual
%% Note that you should use the \pkg{}, \proglang{} and \code{} commands.

\section{Introduction}

Modeling binary and categorical data is one of the most commonly encountered tasks of applied statisticians and econometricians. Binary probit and logit models, as well as their extensions to multinomial and binomial outcomes, are widely used. In this vignette, we present \pkg{UPG}, an \proglang{R} package for Bayesian analysis of well-known binary and categorical data models. \pkg{UPG} is based on a number of highly efficient '\textbf{U}ltimate \textbf{P}{\'o}lya \textbf{G}amma' Markov chain Monte Carlo (MCMC) algorithms that have been developed in \citet{fruhwirth2020ultimate}. The package also features a number of 'plug\&play' solutions to facilitate analysis and communication of results.

\pkg{UPG} is especially well suited for analysis of imbalanced data, as the implemented algorithms make efficient posterior simulation possible in these settings. Bayesian analysis of imbalanced data has so far not been the focus of any package released in \proglang{R} while being a highly relevant problem in applied statistics (\citealp{johndrow2019mcmc, fruhwirth2020ultimate}). In general, the Bayesian paradigm has a number of pronounced benefits when it comes to estimation of binary and categorical data regression models. Besides the intuitive appeal of Bayesian uncertainty quantification, it is well known that Bayesian methods are useful in situations characterized by \textit{perfect separation}. This phenomenon occurs when a given covariate (quasi-)perfectly separates the outcome variable of interest. To avoid that parameters drift off to $\pm \infty$ in such scenarios, frequentist statistics suggests for instance penalized likelihood methods (\citealp{heinze2002solution}). In a Bayesian context, the combination of a potentially ill-defined likelihood function with a weakly informative prior with finite support usually suffices to resolve the issues related to perfect separation \citep{gelman2008weakly, rainey2016dealing}. Tightly related to the occurrence of perfect separation are scenarios where certain outcome categories are observed only very rarely or not at all. For similar reasons, Bayesian inference is often able to avoid implausible parameter estimates in these cases. %In fact, \pkg{UPG} is especially well suited for estimation in such imbalanced data scenarios, as the implemented algorithms will offer a high level of efficiency even for extremely imbalanced data sets (\citealp{fruhwirth2020ultimate}).

Apart from the practical and methodological benefits raised above, \pkg{UPG} aims to provide a range of functionality in order to be appealing to different groups of \proglang{R} users. First, researchers that are already familiar with Bayesian statistical analysis can easily introduce the underlying MCMC algorithms in \pkg{UPG} as an additional sampling block to pre-existing Gibbs sampling algorithms using a few lines of code. This may prove useful in several applications, including mixture-of-experts models (\citealp{gormley2019mixture}) or analysis of Markov switching models (\citealp{fruhwirth2006finite}). Second, for a much broader group of users, the package implements methods for easy and fast production of tables and plots from the estimation output provided. This also facilitates analysis for users that are not commonly working within the Bayesian paradigm. 

\pkg{UPG} is licensed under the GNU General Public License 3 and is openly available on the Comprehensive \proglang{R} Archive Network (CRAN, https://cran.r-project.org/package=UPG).

The remainder of this article is structured as follows. Section 2 provides a short overview of the methodology behind \pkg{UPG}. Section 3 gives a brief introduction to the package, intended as a quick-start guide. Section 4 presents an extended illustration of the functionality of \pkg{UPG}. Finally, Section 5 concludes.

\section{Brief methodological overview}

This section provides a very brief summary of the latent representations underlying the inner workings of the models implemented in \pkg{UPG}. Most of the contents and ideas are directly taken from  \citet{fruhwirth2020ultimate}, where the authors develop the methodology underlying \pkg{UPG}. This is also where the reader is referred to for full theoretical and computational details.

\subsection{Binary regression}

Binary regression models for a set of $N$ binary data $\ym=(y_1, \ldots,y_N)$ are defined by
\begin{equation}\label{eq:binm}
  \Probab{y_i=1|\Xbeta_{i}, \betav}= \Ferror (\Xbeta_{i} \betav).
  \end{equation}
   Choosing  the cdf $\Ferror(\errordiff)=\Normalcdf (\errordiff)$ of the standard normal distribution
 leads to the probit model $\Probab{y_i=1|\Xbeta_{i},\betav} = \Normalcdf( \Xbeta_{i} \betav)$, whereas
  the cdf $\Ferror(\errordiff)=e^\errordiff/(1+e^\errordiff)$ of the logistic distribution
leads to the  logit model $$\Probab{y_i=1| \Xbeta_{i},\betav}=  e^{\Xbeta_{i} \betav}/(1+  e^{\Xbeta_{i} \betav}).$$

A latent variable representation of model (\ref{eq:binm}) involving the latent utility  $\yprodiff_{i}$ is given by:
 \begin{eqnarray}
y_i=\indic{ \yprodiff_{i}>0}, && \displaystyle   \yprodiff_{i} =  \Xbeta_{i} \betav + \errordiff_{i},
 \quad  \errordiff_{i} \sim \ferror(\errordiff_i),\label{eq:binlat}
\end{eqnarray}
where  $\ferror(\errordiff)= \Ferror' (\errordiff) = \Normalpdf (\errordiff)$ is equal to the standard normal  pdf for a probit model and equal to $\ferror(\errordiff)= e^{\errordiff}/(1+e^{\errordiff})^2$ for a logit model.

While MCMC estimation based on (\ref{eq:binlat}) is straightforward for the probit model using one level of data augmentation
 involving the latent utilities $(\yprodiff_{1}, \ldots, \yprodiff_{N})$ \citep{alb-chi:bay_ana},
for the logit model a second level of data augmentation is required in addition to $\yprodiff_{i}$, based  on a
mixture representation of the logistic distribution.
In \pkg{UPG}, we apply the mixture representation of the logistic distribution also used in \citet{fruhwirth2020ultimate},
  \begin{eqnarray}  \label{PGHW}
   \ferror(\errordiff)= e^{\errordiff}/(1+e^{\errordiff})^2 = \frac{1}{4}\int  e ^{- \omegaH \, \errordiff^2/2} p(\omegaH) d\,\omegaH ,
    \end{eqnarray}
   where   $\omegaH \sim \PG{2,0}$ follows a  P\'{o}lya-Gamma distribution \citep{pol-etal:bay_inf}
   with parameters $b=2$ and $\kappa=0$. Conveniently, $\omegaH_i | \errordiff_i$ again follows a P\'{o}lya-Gamma distribution which is easy to sample from. This allows to set up a Gibbs sampler for posterior simulation in a rather straightforward manner.

\subsection{Multinomial logistic regression}

Let  $\left\{y_{i}\right\}$ be a sequence of categorical data,
$i=1,\dots, N$, where  $y_{i}$ is equal to one of at least three unordered
categories. The categories are  labeled by $\labset=\{0,\dots,m\}$,
and for any $k$ the set of all categories but $k$ is denoted by
$\labset_{-k}= L  \setminus \{k\}$.
 We assume that the observations are mutually independent and that for each $k \in L$ the
probability of $y_{i}$ taking the value $k$ depends on covariates $\Xbeta_{i}$   in the following way:
 \begin{eqnarray} \label{mraglam}
\Prob{(y_{i}=k |\betav_0,\dots,\betav_m)}% |\pltm{k}{it}}%
=\pl_{ki} (\betav_0,\dots,\betav_m)  %\label{mragmm} %\\ && \pltm{k}{it}  =
= \displaystyle  \frac{ \exp(\Xbeta_{i} \betav_{k})}
 {\displaystyle\sum_{l=0}^m \exp(\Xbeta_{i}\betav_{l})},
 \end{eqnarray}
 where $\betav_0, \dots, \betav_m$ are category specific  unknown
 parameters of dimension $\betad$.  To make the model identifiable, the parameter $\betav_{k_0}$ of a baseline
category $k_0$ is set equal to $\bfz$: $\betav_{k_0}=\bfz$. Thus, the
parameter $\betav_k$ is relative to the baseline category $k_0$ in terms of the change in
log-odds. In the following, we assume without
loss of generality that $k_0=0$.
This multinomial regression model has the following well-known representation:
\begin{eqnarray} \label{pdlatTR} 
&& \yprodiff_{ki}= \Xbeta_i \betav_k - \xi_{ki} (\betav_{-k})  + \errordiff_{ki}, \quad 
\errordiff_{ki} \sim  \Logistic\\ \label{pdchoice}
&& \displaystyle y_{i }= 
\left\{ \begin{array}{ll}
k, &   \yprodiff_{ki} > 0,  \\
\neq k, &  \yprodiff_{ki} \leq 0.
\end{array} \right. \label{locTR}
\end{eqnarray}
where the error term $\errordiff_{ki}$  follows a logistic distribution,  $\yprodiff _{ki} =  \ypro_{ki} - \max_{\ell \in \labset_{-k}} \ypro_{\ell,i}$ is the utility gap between category $k$ and all its alternatives. %i.e.:
%\footnote{\SFS{Note that in Appendix~\ref{sec:details_mnl_tria} the notation $d_{ki}$ has been used to denote the utility gap.}}
%\begin{eqnarray*}
%\yprodiff _{ki} =  \ypro_{ki} - \max_{\ell \labset_{-k}} \ypro_{\ell,i}, 
%\end{eqnarray*}
and the offset $\xi_{ki} (\betav_{-k})$  is defined as:
\begin{eqnarray*}
&& \xi_{ki} (\betav_{-k})= \log \left(1+ \sum_{\ell \neq \{k,0\}} \exp ( \Xbeta_i \betav_\ell)\right).
\end{eqnarray*}

\pkg{UPG} uses a Gibbs sampling scheme based on this representation. The details of the scheme and the underlying MCMC boosting algorithm are given in \citet{fruhwirth2020ultimate}.

\subsection{Binomial logistic regression}

Finally, \pkg{UPG} can handle regression models with binomial outcomes, i.e. models of the form
     \begin{eqnarray}  \label{obsljuhbin}
 y_i \sim \Bino{N_i, \pi_i}, \qquad  \logit \, \pi_i= \Xbeta_{i} \betav, \qquad i=1,\ldots, N,
 \end{eqnarray}
where $y_i$ can be interpreted as the number of successes out of $N_i$ trials of individual $i$. As shown in \citet{fruhwirth2020ultimate},  the binomial model has the following  random utility  representation  for $0 < y_i<N_i$:
\begin{eqnarray} \label{aggnew3U}
&& \yprow_{i} =  \Xbeta_{i} \betav  +   \errordiff_{\yprow ,i},  \quad   \errordiff_{\yprow ,i} \sim {\GenLogistic{II}{k}}, \\
&& \yprov_{i}  =  \Xbeta_{i} \betav  +   \errordiff_{\yprov ,i},  \quad     \errordiff_{\yprov ,i} \sim  {\GenLogistic{I}{N_i-k}}, \nonumber \\
&&  y_i=k  \Leftrightarrow   \yprow_{i}>0,  \yprov_{i} <0,  \nonumber
\end{eqnarray}
where  $\GenLogistic{I}{\nu}$ and $\GenLogistic{II}{\nu}$ are,  respectively,
the generalized logistic distributions of  type I and type II.
% \item
For  $y_i=0$, the  model  reduces to
\begin{eqnarray*} %\label{aggnew3aU}
\yprov_{i}   =  \Xbeta_{i} \betav  +   \errordiff_{\yprov ,i},  \quad     \errordiff_{\yprov ,i} \sim  {\GenLogistic{I}{N_i}} , \quad  y_i=0 \Leftrightarrow  \yprov_{i}  <0.  \nonumber
\end{eqnarray*}
%\item
For  $y_i=N_i$, the  model  reduces to
\begin{eqnarray*}% \label{aggnew3bU}
  \yprow_{i}=  \Xbeta_{i} \betav  +   \errordiff_{\yprow ,i},  \quad   \errordiff_{\yprow ,i} \sim  {\GenLogistic{II}{N_i}}, \quad  y_i=N_i \Leftrightarrow
  \yprow_{i} >0.  \nonumber
\end{eqnarray*}
%\end{enumerate}

\noindent
For $N_i=1$, the logistic  model results, as both  $\GenLogistic{I}{\nu}$ and $\GenLogistic{II}{\nu}$ reduce to a
 logistic distribution for $\nu=1$. For  $y_i=0$, $z_i = \yprov_{i}$,
whereas  for $y_i=1$, $z_i= \yprow_{i}$, and the choice equation reduces to $y_i=\indic{z_i>0}$. To estimate $\betav$ in this framework, it is possible to derive mixture representations similar to (\ref{PGHW}) for the $\GenLogistic{I}{\nu}$ and $\GenLogistic{II}{\nu}$ error distributions, see \citet{fruhwirth2020ultimate} for details.

\subsection{Increasing sampling efficiency through marginal data augmentation}
\label{sec:boosting}

It is well known that Bayesian estimation of binary and categorical data models using data augmentation may result in inefficient sampling behavior, especially in settings with imbalanced data (\citealp{johndrow2019mcmc}). The samplers that are outlined in the previous subsections are, in principle, no exemption from this rule. To tackle this issue, \pkg{UPG} implements \textit{boosted} MCMC algorithms that have been developed in \citet{fruhwirth2020ultimate} to enable highly efficient posterior sampling in a broad range of settings. These MCMC boosting methods are similar in spirit to previous work on MCMC sampling efficiency, see for instance \citet{kastner2014ancillarity} or \citet{kastner2017efficient} for MCMC boosting in the context of (factor) stochastic volatility models. Specifically, the sampling algorithms available in \pkg{UPG} rely on marginal data augmentation \citep{liu:wu, van-men:art} to increase sampling efficiency. This involves location-based and scale-based expansion of the latent variable representations introduced above. Theoretical and computational details, as well as a number of large-scale simulation studies demonstrating the potential gains in sampling efficiency, may be found in \citet{fruhwirth2020ultimate}.

\section{UPG Basics}

The \pkg{UPG} package provides efficient sampling algorithms for Bayesian analysis of the probit, logit, multinomial logit and binomial logit model. This section covers the basics of the package, including data requirements, estimation as well as the methods included in \pkg{UPG}.

In terms of inputs, the minimum requirement for probit, logit and multinomial logit models is a suitable $N \times 1$ dependent vector \code{y} and a $N \times d$ design matrix \code{X}. An additional $N \times 1$ vector of total number of trials \code{Ni} is necessary to estimate a binomial logit model. For probit and logit models, \code{y} is supposed to be binary. For multinomial logit models, \code{y} is a categorical vector containing one realized category out of the set $L = \{0, \ldots, m\}$ for each observation. The baseline category $k_0$ can be freely chosen by the user through parameter \code{baseline}. If no baseline is provided, the most frequently observed category is used as baseline. For binomial logits, \code{y} contains the number of successes of each observation. Inputs of class \code{integer}, \code{numeric}, \code{matrix} and \code{data.frame} are accepted. In the multinomial logit case, \code{character} and \code{factor} are accepted as dependent vector types as well. Depending on the specified model type, \pkg{UPG} will use a variety of data checks to ensure proper estimation. 

The necessary tools for efficient estimation of binary and categorical data models in a Gibbs sampling framework are wrapped into a single estimation function \code{UPG()} to provide a minimalistic user interface. The four different models included in \pkg{UPG} can be called using the \code{model} parameter as shown in \autoref{tab:estimation}. An illustration of the estimation process and the most important posterior analysis methods using \pkg{UPG} are discussed in the next section.

\begin{table}[t]
    \centering
    \begin{tabular}{ll}
     \textbf{Estimation Command}    & \textbf{Model} \\
    \code{UPG(y, X, model = "probit")}     & Probit \\
        \code{UPG(y, X, model = "logit")}     & Logit \\
    \code{UPG(y, X, model = "mnl")}     & Multinomial Logit \\
    \code{UPG(y, X, Ni, model = "binomial")}     & Binomial Logit \\

    \end{tabular}
    \caption{Estimation commands for the models included in \pkg{UPG}}
    \label{tab:estimation}
\end{table}

In terms of output, \code{UPG()} will return one out of four S3 object classes, depending on the specified \code{model}. The classes are \code{UPG.Probit}, \code{UPG.Logit}, \code{UPG.MNL} and \code{UPG.Binomial}. These objects hold the full posterior distribution for all parameters. In addition, all user inputs are copied into the output object for further analysis. Several \code{S3} methods can be applied to any of these objects. The main task of these methods is to conveniently summarize the generated posterior samples. The methods themselves are summarized in \autoref{tab:functions} and will be discussed in further detail in the subsequent section using extensive examples.

\begin{table}[t]
    \centering
    \begin{tabular}{ll}
    \textbf{S3 Method}    & \textbf{Usage} \\
    \code{print}       & Object overview \\
    \code{summary}     & Summary of posterior estimates as table output \\
    \code{plot}        & Plot coefficient point estimates and credible intervals \\
    \code{predict}     & Predict probabilities for new data or input data \\
    \code{coef} & Extract posterior means and credible intervals of coefficients \\
    \code{logLik} & Extract log-likelihood based on posterior mean
    \end{tabular}
    \caption{\code{S3} methods included in \pkg{UPG}}
    \label{tab:functions}
\end{table}

\section{Analyzing binary and categorical data using UPG}

In this section, the included data sets in the \pkg{UPG} are presented and utilized as sample cases to showcase the package's capabilities.

\subsection{Bayesian binary regression: Probit and Logit}

To demonstrate how to estimate and analyze Bayesian probit and logit models using \pkg{UPG}, a microeconomic data set on female labor force participation from the US \textit{Panel Study of Income Dynamics} is included. It features a binary variable indicating labor force status as well as a number of additional covariates for 753 women:
\begin{Sinput}
R>  data("lfp", package = "UPG")
R>  head(lfp, 5)
\end{Sinput}

\begin{Soutput}
  lfp intercept k5 k618        age wc hc       lwg   inc
1   1         1  1    0 -1.3053889  0  0 1.2101647 10.91
2   1         1  0    2 -1.5531414  0  0 0.3285041 19.50
3   1         1  1    3 -0.9337602  0  0 1.5141279 12.04
4   1         1  0    3 -1.0576365  0  0 0.0921151  6.80
5   1         1  1    2 -1.4292651  1  0 1.5242802 20.10
\end{Soutput}

The binary dependent variable \code{lfp} takes the value of $1$ if the woman is participating in the labor force. \code{k5} gives the number of children under the age of 5, \code{k618} indicates the number of children between 6 and 18 years, \code{age} is a standardized age index and \code{wc} as well as \code{hc} are binary indicators capturing whether a college degree was obtained by the wife and the husband, respectively. In addition, two income related predictors are included, where \code{lwg} describes the expected log wage of the woman and \code{inc} gives the logarithm of family income exclusive of the income of the woman. This data set comes from the \pkg{carData} package and has been originally analyzed in \citet{mroz1987sensitivity}.

\subsubsection{Model estimation}

To construct a suitable design matrix \code{X} and a binary dependent vector \code{y} for probit and logit models, it suffices to split the data set as follows:

\begin{Sinput}
R>  y <- lfp[, 1]
R>  X <- lfp[, -1]
\end{Sinput}

In order to estimate a Bayesian logit model, we can use

\begin{Sinput}
R>  results.logit <- UPG(y = y, X = X, model = "logit")
\end{Sinput}
\begin{Soutput}
Checking data & inputs ... 
Initializing Gibbs Sampler ... 
Simulating from posterior distribution ... 
  |=============================================| 100%
Sampling succesful!
Saving output ...
Finished! Posterior simulation took 2.8 seconds. 
\end{Soutput}

In the remainder of this subsection, it is assumed that the goal is to estimate and analyze a logit model using \code{model = 'logit'}. Changing the \code{type} parameter to \code{model = 'probit'} allows to estimate a probit model. The syntax in the illustration below holds for both models. 

\subsubsection{Tabulating results}

Applying \code{summary} to the output object results in a quick overview of the regression results in the form of tabulated parameter estimates. Continuing the running example, it is easy to generate a table with posterior means and standard deviations as well as credible intervals:

\begin{Sinput}
R>  summary(results.logit)
\end{Sinput}

\begin{Soutput}
--- Bayesian Logit Results --- 

N = 753 
Analysis based on 1000 posterior draws after
an initial burn-in period of 1000 iterations.
MCMC sampling took a total of 2.8 seconds. 


|          |  Mean|   SD|  Q2.5| Q97.5| 95% CI excl. 0 |
|:---------|-----:|----:|-----:|-----:|:--------------:|
|intercept |  0.50| 0.24|  0.05|  1.00|       *        |
|k5        | -1.47| 0.20| -1.85| -1.11|       *        |
|k618      | -0.06| 0.07| -0.20|  0.07|                |
|age       | -0.50| 0.11| -0.72| -0.31|       *        |
|wc        |  0.81| 0.24|  0.34|  1.26|       *        |
|hc        |  0.12| 0.21| -0.27|  0.52|                |
|lwg       |  0.61| 0.15|  0.31|  0.92|       *        |
|inc       | -0.04| 0.01| -0.05| -0.02|       *        |
\end{Soutput}

In terms of interpretation, it is for instance visible that women with a college degree (\code{wc}) are more likely to participate in the labor force compared to women with no formal tertiary education, holding everything else constant. On the contrary, women who have small children under the age of 5 (\code{k5}) are \textit{ceteris paribus} less likely to be active in the labor force compared to women without young children.

A number of possibilities for exporting summary tables to \LaTeX, HTML or Microsoft Word using \code{summary(obj, type = c("html","latex","pandoc"))} are available.\footnote{If a LaTeX table is desired, the 'booktabs' package has to be included in the preambel of the LaTeX document.} The user can choose from a number of different options to customize the table output directly, including credible intervals based on posterior quantiles specified using \code{q}, the names of the variables using \code{names}, the number of significant digits using \code{digits}, the subset of variables to be used in the table using \code{include} and the table caption using \code{cap}. Further customizations are easy to implement, as \code{summary} returns a \code{knitr_kable} object that can be further modified using the \pkg{knitr} package (\citealp{knitr_package}). More details can be found in the package manual.

\subsubsection{Visualizing results}

In case a more visual representation of the model output is desired, the \code{plot} function can be used to generate publication-ready coefficient plots for all four available models using \pkg{ggplot2} (\citealp{ggplot2}). Similar to the \code{summary} function, \code{plot} allows the user to customize a number of pre-specified parameters such as axis labels (\code{xlab}, \code{ylab}), coefficient names (\code{names}), the reported posterior quantiles (\code{q}), and the set of included variables (\code{include}). \code{plot} will return a \code{ggplot2} object that can be further modified using the arsenal of tools from the \code{ggplot2} universe.

Continuing the logit example, we can generate a simple coefficient plot using

\begin{Sinput}
R>  plot(results.logit)
\end{Sinput}

\begin{center}
        \includegraphics[width=.6\textwidth]{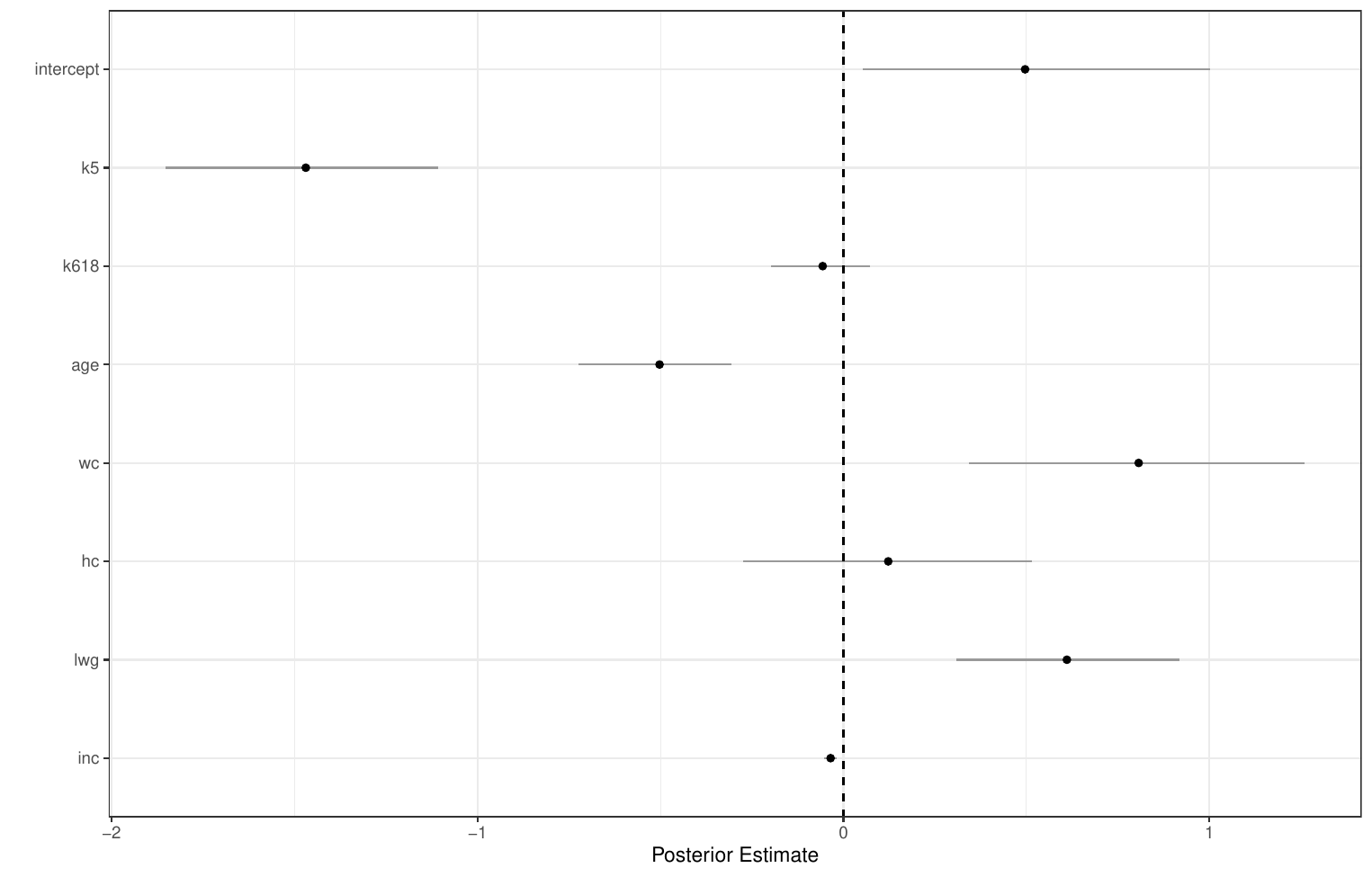}
\end{center}

These plots provide point estimates as well as credible intervals for each covariate by default. The variables may be sorted by estimated effect size using \code{sort = TRUE}. Otherwise, they appear in the same order as in \code{X}.

\subsubsection{Predicting probabilities}

In several situations, applied researchers are not necessarily interested in examining the estimated coefficients, but in using these estimates to generate predictions. For these scenarios, \code{predict} may be used to produce point predictions and prediction intervals based on the estimated model. These predictions can be generated using either the data provided for model estimation or new, external data provided by the user. Continuing the running example, 

\begin{Sinput}
R>  predict(results.logit)
\end{Sinput}

will return a list containing the posterior mean as well as the 97.5\% and 2.5\% posterior quantiles of the posterior predictive distribution given the input data. In case the user wants to predict probabilities using external data, a suitable explanatory matrix with the same number of columns and same variable ordering must be provided. The syntax in that case is

\begin{Sinput}
R>  predict(results.logit, newdata = X.new)
\end{Sinput}

where \code{X.new} is the new design matrix used for prediction. Similar to the other available \code{S3} methods in \pkg{UPG}, the reported posterior quantiles can be specified using the parameter \code{q}.

\subsubsection{Log-likelihood}

In case the user is interested in the log-likelihood of the data given the parameters, a \code{logLik} method is available. Applying this method to the output will extract the log-likelihood evaluated at the posterior mean of the parameters:

\begin{Sinput}
R>  logLik(results.logit)
\end{Sinput}
\begin{Soutput}
'log Lik.' -452.645 (df=8)
\end{Soutput}

This log-likelihood object also holds information on the number of observations as well as the number of estimated parameters. 

% is constructed to be compatible with commonly used model selection criteria functions such as \code{AIC} and \code{BIC}.

% \footnote{Note that the AIC assumes that the estimates are based on maximum likelihood estimation and should thus be used with caution for model selection using \pkg{UPG}.}

\subsection{Bayesian binomial logistic regression}

To demonstrate how to estimate a binomial logit model using \pkg{UPG}, aggregated individual passenger data of the \textit{RMS Titanic} is included as an example data set:

\begin{Sinput}
R>  data("titanic", package = "UPG")
R>  head(titanic, 5)
\end{Sinput}
\begin{Soutput}
  survived total intercept pclass female age.group
1        0     1         1      1      1         5
2        5     5         1      2      1         5
3       12    17         1      3      1         5
4        2     2         1      1      0         5
5        8     8         1      2      0         5
\end{Soutput}

The passengers have been split into several groups based on passenger class (\code{pclass}), five year age groups (\code{age.group}) and gender (\code{female}). For each group, total passenger counts (\code{total}) and the number of passengers that survived the disaster (\code{survived}) are provided. The data set is an aggregate version of the well-known titanic data set (\citealp{hilbe2007negative}, Table 6.11) that has for instance been previously analyzed in \citet{fruehwirth2009improved}.\footnote{See \url{https://www.kaggle.com/c/titanic/} for more details.}

\subsubsection{Model estimation}

In this case, the dependent vector of successes is \code{survived} whereas the number of total trials corresponds to \code{total}. Both vectors have to be provided in addition to explanatory variables to be able to estimate a binomial logit model using \pkg{UPG}. Hence, the data needs to be split into three parts parts prior to estimation:

\begin{Sinput}
R>  y  <- titanic[,1]
R>  Ni <- titanic[,2]
R>  X  <- titanic[,-c(1,2)]
R>  results.binomial <- UPG(y = y, X = X, Ni = Ni, model = "binomial")
\end{Sinput}
\begin{Soutput}
Checking data & inputs ... 
Initializing Gibbs Sampler ... 
Simulating from posterior distribution ... 
  |=============================================| 100%
Sampling succesful!
Saving output ...
Finished! Posterior simulation took 1.84 seconds. 
\end{Soutput}

All further steps of analysis are similar to the binary logit regression example. As an example, we tabulate the results using a credible interval based on the 10\% and 90\% posterior quantiles:

\begin{Sinput}
R>  summary(results.binomial, q = c(0.1, 0.9))
\end{Sinput}
\begin{Soutput}
--- Bayesian Binomial Logit Results --- 

N = 78 
Analysis based on 1000 posterior draws after
an initial burn-in period of 1000 iterations.
MCMC sampling took a total of 1.84 seconds. 


|          |  Mean|   SD|   Q10|   Q90| 80% CI excl. 0 |
|:---------|-----:|----:|-----:|-----:|:--------------:|
|intercept |  2.23| 0.43|  1.68|  2.80|       *        |
|pclass    | -1.19| 0.12| -1.35| -1.05|       *        |
|female    |  2.58| 0.18|  2.35|  2.81|       *        |
|age.group | -0.03| 0.01| -0.04| -0.02|       *        |
\end{Soutput}

In terms of interpretation, we can for instance see that female passengers have had a much higher survival probability compared to their male counterparts. A higher passenger class (corresponding to cheaper tickets) is associated with higher mortality. Finally, the log-odds of survival decrease with increasing age. 
To demonstrate custom credible intervals when plotting results, the estimation output is visualized using \code{q = c(0.1, 0.9)}. This results in a 80\% credible interval based on the 0.1 and 0.9 quantiles of the posterior distribution. In addition, custom variable names are provided and \code{sort = TRUE} ensures that the variables are ordered based on estimated (average) effect size:

\begin{Sinput}
R>  plot(results.binomial,
R>       sort  = TRUE,
R>       q     = c(0.1, 0.9),
R>       names = c("Intercept", "Passenger Class", "Female", "Age Group"))
\end{Sinput}

\begin{center}
        \includegraphics[width=.6\textwidth]{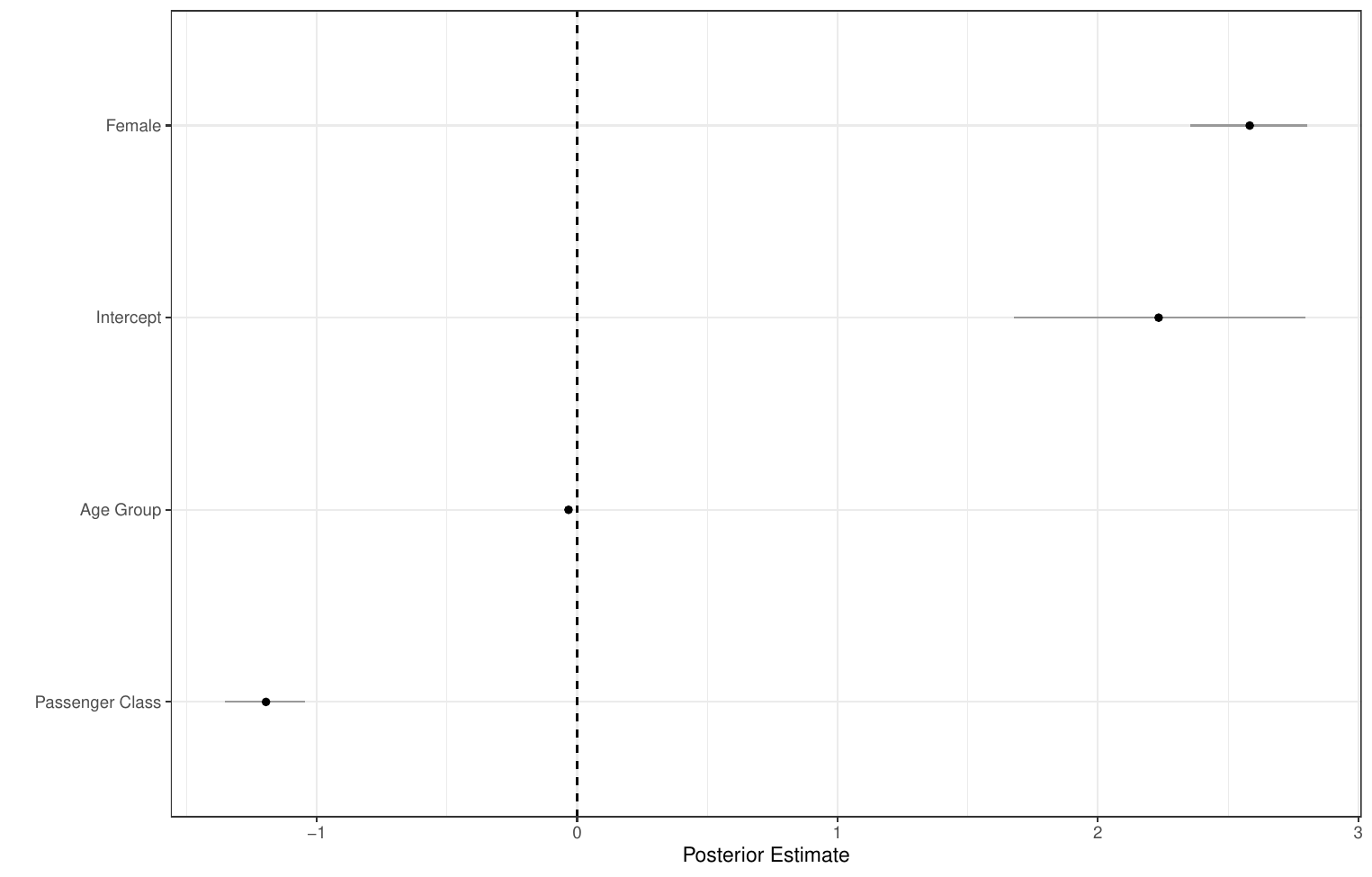}
\end{center}

\subsection{Bayesian multinomial logistic regression}

For the multinomial logit model, a data set on 200 high school students and their program choice (general, vocational or academic) is included together with a binary variable taking the value of 1 for female students (\code{female}), a categorical variable indicating socio-economic status (\code{ses}) and standardized results of a writing test (\code{write}):

\begin{Sinput}
R>  data("program",package="UPG")
R>  head(program,5)
\end{Sinput}

\begin{Soutput}
   program intercept female ses     write
1 vocation         1      1   1 -1.875280
2  general         1      0   2 -2.086282
3 vocation         1      0   3 -1.453276
4 vocation         1      0   1 -1.664278
5 vocation         1      0   2 -2.297284
\end{Soutput}

This data set is also known as the \code{hsbdemo} data set and is provided online by the University of California, Los Angeles Statistical Consulting Group. This data is used in several R packages and in other software tools as example data for multinomial logistic regression.\footnote{See for instance \url{https://stats.idre.ucla.edu/stata/dae/multinomiallogistic-regression/} for usage of the data in Stata.}

\subsubsection{Model estimation}

As mentioned above, dependent variables for multinomial logit estimation have to be provided as a categorical vector. By default, the category that occurs most often is chosen as baseline category. An alternative baseline category may be specified using \code{baseline}. In the example data set, \code{academic} is chosen 105 times out of 200 observations and will thus serve as baseline category. The code to create \code{y} and \code{X} is quite similar to the probit and logit case:

\begin{Sinput}
R>  y <- program[,1]
R>  X <- program[,-1]
\end{Sinput}

To estimate a multinomial logit model, \code{model = 'mnl'} has to be specified when using the \code{UPG} command:

\begin{Sinput}
R>  results.mnl <- UPG(y = y, X = X, model = 'mnl', verbose = FALSE)
\end{Sinput}

where we have set \code{verbose = FALSE} to suppress any output during estimation for illustration purposes. Handling the resulting \code{UPG.MNL} object is similar to the cases outlined above and is thus only discussed briefly. Tabulation of the results is based on a grouped representation of the model output:

\begin{Sinput}
R>  summary(results.mnl, 
R>          names  = c("Intercept", "Female", "SES", "Writing Score"))
\end{Sinput}

\begin{Soutput}
--- Bayesian Multinomial Logit Results --- 

N = 200 
Analysis based on 1000 posterior draws after
an initial burn-in period of 1000 iterations.
MCMC sampling took a total of 2.35 seconds. 

Category 'academic' is the baseline category.


|                    |  Mean|   SD|  Q2.5| Q97.5| 95% CI excl. 0 |
|:-------------------|-----:|----:|-----:|-----:|:--------------:|
|Category 'general'  |      |     |      |      |                |
|Intercept           |  0.43| 0.58| -0.68|  1.58|                |
|Female              |  0.08| 0.37| -0.69|  0.75|                |
|SES                 | -0.62| 0.26| -1.12| -0.12|       *        |
|Writing Score       | -0.56| 0.22| -0.99| -0.13|       *        |
|                    |      |     |      |      |                |
|Category 'vocation' |      |     |      |      |                |
|Intercept           | -0.32| 0.65| -1.65|  0.92|                |
|Female              |  0.49| 0.40| -0.29|  1.30|                |
|SES                 | -0.40| 0.28| -0.97|  0.12|                |
|Writing Score       | -1.15| 0.21| -1.62| -0.78|       *        |
\end{Soutput}

From the output, it becomes obvious that, in the observed sample, higher scores on a writing test decrease the probability of choosing a general or vocational program compared to the baseline of choosing an academic program. Similar conclusions can be drawn from a coefficient plot that is grouped by outcome category:

\begin{Sinput}
plot(results.mnl, 
     names  = c("Intercept", "Female", "SES", "Writing Score"))
\end{Sinput}

\begin{center}
        \includegraphics[width=.6\textwidth]{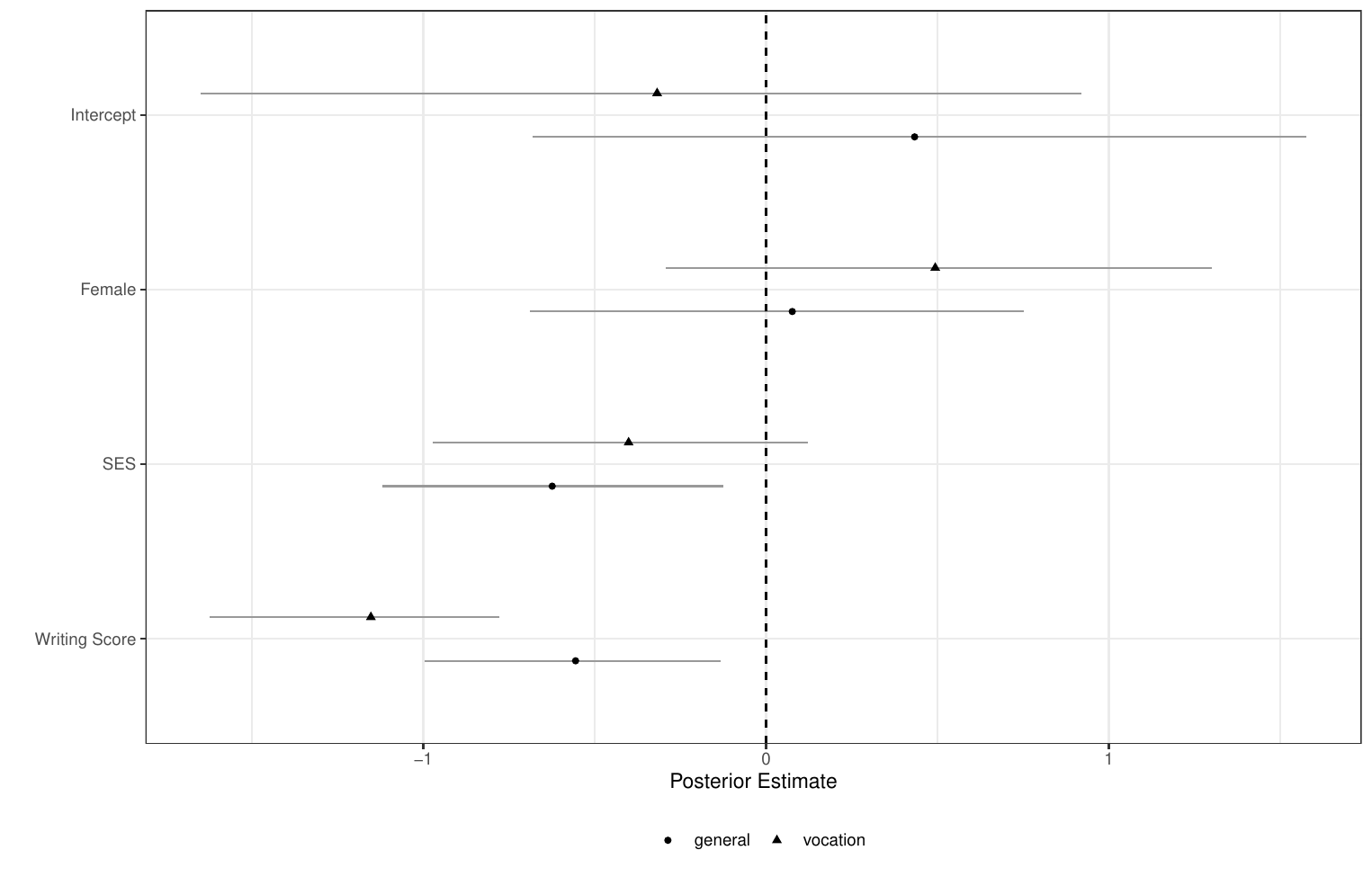}
\end{center}

\subsection{UPG-within-Gibbs}

In certain applications, users might want to use \pkg{UPG} as a single sampling step within a pre-existing Gibbs sampler. Examples where this might be useful include mixture-of-experts models, where a multinomial logit prior can be implemented (see e.g. \citealp{gormley2019mixture}). Similarly, probits as well as binary and multinomial logits do often serve as prior models in Bayesian Markov switching frameworks (\citealp{fruhwirth2006finite}).

To implement '\code{UPG}-within-Gibbs' it is possible to access the underlying sampling algorithms directly in order to bypass all checks. Consider the example of a binary logit model. Assuming a starting value for \code{beta.draw} is given, it suffices to add

\begin{Sinput}
draw      = UPG:::upg.logit(y, 
                            X, 
                            nsave = 1, 
                            nburn = 0, 
                            verbose = F, 
                            beta.start = beta.draw)  
beta.draw = t(draw$beta)  
\end{Sinput}

as code block in an existing Gibbs sampler. In this example, \code{nsave} is set to $1$ and \code{nburn} is set to 0 to generate exactly one posterior sample without burn-in period. \code{verbose = F} suppresses all console output and parameter \code{beta.start} is used to provide the current value of \code{beta.draw} as starting value. Iterating over this code $M$ times and saving the resulting posterior draws of \code{beta} gives equivalent results to generating $M$ posterior draws from \code{UPG} directly. However, note that this will, in general, be slightly slower than generating $M$ samples from \code{UPG} directly. This is due to the overhead that results from repeated function calls in R. %Specifically, generating $M$ posterior samples from \code{UPG} directly results in exactly one function call, while '\code{UPG}-within-Gibbs' will amount to $M$ function calls. However, even in this 'slow' scenario, \pkg{UPG} allows for rather fast sampling. For instance, the 'within-Gibbs' logistic regression model outlined above samples more than 270 posterior draws per second using the example data with $N = 753$ and eight covariates on an Intel i7 CPU @ 2.40GHz. A more detailed picture on sampling efficiency and sampling speed is provided in the next subsection.

\subsection{Further details}

The estimation of binary, multinomial and binomial logit models requires simulating from P{\'o}lya Gamma distributions. This is accomplished using an implementation in \pkg{pgdraw} (\citealp{pgdraw}). In terms of prior distributions, the elements of $\betav$ are assumed to follow independent Gaussian distributions a priori. In order to change the prior variances, the parameters \code{A0} and \code{B0} of the \code{UPG} function are available, refering to the variances of the intercept and the remaining coefficients, respectively. Both \text{A0} and \text{B0} have 4 as default value.

\subsection{Sampling efficiency and sampling speed}

In order to shed some light on the performance of the implemented models in specific applications, the user can compute several MCMC diagnostic measures using the command \code{UPG.Diag}. Specifically, a call to \code{UPG.Diag} will return the effective sample size (ESS) for each coefficient derived using \code{effectiveSize} from \pkg{coda} (\citealp{coda}).\footnote{Effective sample sizes in \pkg{coda} are derived from the spectral density at 0 which is estimated based on fitting an autoregressive process to the posterior draws.} In addition, inefficiency factors (IE; given by the number of saved draws divided by the effective sample size) and the effective sampling rate (ESR; given by the effective sample size divided by the running time of the sampler in seconds) are returned. To allow for a more convenient 'quick check' of the behavior of the Markov chain, \code{UPG.Diag} also returns the minimum, maximum and median across all coefficients as summary statistics of these three diagnostic measures. 

To give a sense of magnitude with respect to computational performance, we summarize ESS, IE and ESR for probit as well as binary, binomial and multinomial logit models using the presented example data sets. For each model, 10,000 posterior draws are sampled after an initial burn-in period of 1,000 iterations. All simulations have been run on an AMD Ryzen 5 5500U. The results of this exercise are shown in \autoref{tab:speed}. While the table shows that the MCMC algorithms in \pkg{UPG} exhibit rather efficient sampling behavior, a pronounced drop in sampling \textit{speed} is visible when comparing the probit regression framework to the remaining models. This is due to the increased computational effort that results from sampling P{\'o}lya Gamma random variables. While these are not needed in the MCMC scheme of the probit model, they are necessary for all logit models in \pkg{UPG}, increasing computation time in each sweep of the sampler. Nevertheless, due to high levels of sampling \textit{efficiency}, an effective posterior sample size that is sufficient for inference can be generated speedily in the logit frameworks as well.% In general, adding a few thousand more effective posterior samples in medium-sized data sets will be possible in significantly less than a minute.

\begin{table}[t]
\centering
\ \\
\begin{tabular}{@{}llrrrr@{}}
\toprule
    &        & Probit & Binary Logit & Multinomial Logit & Binomial Logit \\ \midrule
    & N      & 753    & 753          & 200               & 78             \\
    & d      & 8      & 8            & 4                 & 4              \\
    & $\sum_i N_i$   &       &             &                  & 887            \\\addlinespace
ESS & Min. & 3557 & 2516 & 1507 & 3421\\
 & Median & 3834 & 3055 & 1944 & 3595\\
 & Max. & 3976 & 3193 & 2277 & 3923\\\addlinespace
IE & Min. & 2.51 & 3.13 & 4.39 & 2.55\\
 & Median & 2.61 & 3.27 & 5.14 & 2.78\\
 & Max. & 2.81 & 3.97 & 6.64 & 2.92\\\addlinespace
ESR & Min. & 843 & 200 & 114 & 386\\
 & Median & 909 & 243 & 147 & 406\\
 & Max. & 942 & 254 & 172 & 443\\\addlinespace
Time (in sec.) & & 4.22 & 12.56 & 13.23 & 8.86 \\  \bottomrule
\end{tabular}
\caption{Sampling efficiency and sampling speed of the implemented models. Results are based on 10,000 saved draws after an initial burn-in period of 1,000 iterations using the example data sets as input.}
\label{tab:speed}
\end{table}

\section{Conclusion}

In this vignette, the \proglang{R} package \pkg{UPG} is introduced as a software tool for Bayesian estimation of probit, logit, multinomial and binomial logit models. In addition to an implementation that enables efficient estimation through marginal data augmentation, the package is designed to provide easy access to Bayesian models for binary and categorical data for researchers that might not be familiar with the Bayesian paradigm. At the same time, users have the possibility to easily include the provided models as a new sampling step in existing Gibbs samplers. Moreover, the package includes a variety of functions that may be used to produce tables and plots that summarize the estimation output. These methods have been introduced and illustrated through applied examples using data sets that come with the package.

\section*{Acknowledgments}

The authors would like to thank Maximilian B{\"o}ck, Nikolas Kuschnig, Darjus Hosszejni and Peter Knaus for helpful comments and for being valuable discussion partners during package development.

\small{\addcontentsline{toc}{section}{References}
\bibliography{jss}}

\end{document}